\documentclass[a4paper,11pt]{article}
\usepackage{jcappub} % for details on the use of the package, please
\usepackage[T1]{fontenc} % if needed
\usepackage{bm}

\newcommand{\ms}{M_\odot}
\newcommand{\bmt}{{\bm{\theta}}}

\newcommand{\bmH}{{\bm{H}}}
\newcommand{\rmd}{{\rm{d}}}

\title{Vetting quark-star models with gravitational waves in the hierarchical
Bayesian framework}

\author[a,b]{Ziming Wang,}
\author[c]{Yong Gao,}
\author[d]{Dicong Liang,}
\author[e]{Junjie Zhao}
\author[b,f,1]{and Lijing Shao\note{Corresponding author.}}

\affiliation[a]{Department of Astronomy, School of Physics, Peking University,
Beijing 100871, China}
\affiliation[b]{Kavli Institute for Astronomy and Astrophysics, Peking
University, Beijing 100871, China}
\affiliation[c]{Max Planck Institute for Gravitational Physics (Albert Einstein
Institute), Am M\"uhlenberg 1, D-14476 Potsdam-Golm, Germany}
\affiliation[d]{Department of Mathematics and Physics, School of Biomedical
Engineering, Southern Medical University, Guangzhou 510515, China}
\affiliation[e]{Henan Academy of Sciences, Zhengzhou 450046, Henan, China}
\affiliation[f]{National Astronomical Observatories, Chinese Academy of
Sciences, Beijing 100012, China}

\emailAdd{zwang@pku.edu.cn}
\emailAdd{yong.gao@aei.mpg.de}
\emailAdd{dcliang@smu.edu.cn}
\emailAdd{junjiezhao@hnas.ac.cn}
\emailAdd{lshao@pku.edu.cn}

\abstract{The recent discovery of gravitational waves (GWs) has opened a new
avenue for investigating the equation of state (EOS) of dense matter in compact
stars, which is an outstanding problem in astronomy and nuclear physics. In the
future, next-generation (XG) GW detectors will be constructed, deemed to provide
a large number of high-precision observations.  We investigate the potential of
constraining the EOS of quark stars (QSs) with high-precision measurements of
mass $m$ and tidal deformability $\Lambda$ from the XG GW observatories. We
adopt the widely-used bag model for QSs, consisting of four microscopic
parameters: the effective bag constant $B_{\rm eff}$, the perturbative quantum
chromodynamics correction parameter $a_4$, the strange quark mass $m_s$, and the
pairing energy gap $\Delta$. With the help of hierarchical Bayesian inference,
for the first time we are able to infer the EOS of QSs combining multiple GW
observations.  Using the top 25 loudest GW events in our simulation, we find that,
the constraints on $B_{\rm eff}$ and
$\Delta$ are tightened by several times, while $a_4$ and $m_s$ are still poorly
constrained.  We also study a simplified 2-dimensional (2-d) EOS model which was
recently proposed  in literature.  The 2-d model is found to exhibit significant
parameter-estimation biases as more GW events are analyzed, while the predicted
$m$-$\Lambda$ relation remains consistent with the full model.}

\begin{document}
\maketitle
\flushbottom

%=============================
\section{Introduction}
\label{sec:intro}
%=============================

The state of dense matter in compact stars has been a long-standing  problem in
astronomy and nuclear physics, which originates from the complexity of
nonperturbative quantum chromodynamics (QCD).  Besides the commonly adopted
neutron star (NS) models,  self-bound quark stars (QSs) were also proposed as a
candidate for compact stars.  QSs are entirely occupied by strange quark matter
(SQM), where the SQM forms the true ground state  \cite{Itoh:1970uw,
Bodmer:1971we, Witten:1984rs, Alcock:1986hz, Haensel:1986qb, Alford:2004pf}.
QSs are considered as an important candidate for pulsars, and the properties of
SQM can be constrained by measurements of mass and/or radius in pulsar
observations.  To be consistent with current observation of high-mass pulsars,
one has to take into account the effects from strong interaction, such as
one-gluon exchange or color-superconductivity, to increase the maximum mass of
QSs \cite{Ruester:2003zh, Horvath:2004gn, Alford:2006vz, Fischer:2010wp,
Kurkela:2009gj, Kurkela:2010yk}. Using the theoretically calculated maximum
mass, one can place constraints on the equation of state (EOS) of compact stars,
which gives the relationship between the pressure and the energy density of the
dense matter~\cite{Ozel:2010bz, Lattimer:2010uk, Weissenborn:2011qu,
Bhattacharyya:2016kte, Li:2020wbw, Gao:2021uus}.  Besides the maximum mass, the
radius inferred from electromagnetic observations can provide additional
constraints on the EOS of QSs \cite{Li:2020wbw, daSilva:2023okq}.

Recently, the discovery of gravitational waves
(GWs)~\cite{LIGOScientific:2016aoc} has opened a new window for studying the EOS
of compact stars.  When one compact star orbits around the other in a close
orbit, it gets deformed due to the tidal force exerted by the companion's
gravitational field. Such tidal deformation affects the inspiral of the binary,
thus it is manifested in the phase evolution of GWs~\cite{Vines:2011ud,
Favata:2013rwa, Wade:2014vqa, Baiotti:2019sew, Chatziioannou:2020pqz}.  Given
the mass of a compact star, its tidal deformability, which is a measure of the
star's response to the external gravitational field, is  determined by the EOS.
Consequently, the information about the EOS can be inferred from GW
signals~\cite{Annala:2017tqz, Nandi:2017rhy, Paschalidis:2017qmb, Li:2018ayl,
Gomes:2018eiv, Lai:2018ugk, LIGOScientific:2018cki, Zhang:2018bwq, Zhu:2018ona,
Carney:2018sdv, Tsang:2019vxn, Lim:2019som, GuerraChaves:2019foa, Shao:2022koz,
Ripley:2023lsq}.  Other properties of GW events, such as the oscillation
frequencies of post-merger remnants and the electromagnetic counterparts, can
also provide information on the EOS~\cite{Shibata:2017xdx, Bauswein:2017vtn,
Ruiz:2017due, Shibata:2019ctb, Li:2022qql, Criswell:2022ewn, Zhang:2023zth,
Zhou:2024syq, Li:2024hzt}.

In the future, the next-generation (XG) GW detectors, such as the Cosmic
Explorer (CE)~\cite{Reitze:2019iox, Reitze:2019dyk} and the Einstein Telescope
(ET)~\cite{Punturo:2010zz, Hild:2010id, Sathyaprakash:2012jk} will be
constructed.  Due to their largely increased sensitivity and a lower cutoff
frequency, many more compact binary coalescence signals are expected to be
detected, reaching $10^5$--$10^6$ events per year~\cite{LIGOScientific:2017zlf,
Sathyaprakash:2019yqt, Kalogera:2021bya, Samajdar:2021egv}.  Previous studies
have introduced the hierarchical Bayesian inference techniques to combine
numerous GW events and draw information from populations of GW events
\cite{Mandel:2009pc, Mandel:2009nx, Adams:2012qw, Mandel:2018mve,
Thrane:2018qnx, KAGRA:2021duu}.  In this case, as the EOS determines the
relationship between the tidal deformability $\Lambda$ and the mass $m$ of
compact stars, the EOS parameters can be regarded as hyper-parameters and
incorporated into the framework of hierarchical Bayesian inference.  There are
many studies using this technique to investigate the prospects of exploring the
EOS of NSs in the XG GW detector era, finding that the EOS will be tightly
constrained with the help of  high-precision observation and the accumulation of
GW events~\cite{Lackey:2014fwa, Bose:2017jvk, Landry:2018prl, Talbot:2020oeu,
Golomb:2021tll, Gupta:2022qgg, Criswell:2022ewn, Iacovelli:2023nbv,
Biswas:2023ceq, Walker:2024loo}.

As for the QS model, current studies mostly have used the detected GW events so
far, namely, GW170817, GW190425 and GW190814, to constrain the EOS parameters in
the bag model of QSs~\cite{Zhou:2017pha, Yang:2019rxn, Bombaci:2020vgw,
Zhang:2020jmb, Cao:2020zxi, Lourenco:2021lpn, Miao:2021nuq, Yang:2021sqg,
Yang:2021bpe, Arbanil:2023yil, Oikonomou:2023otn, Podder:2023dey}.  However,
there is still a lack of discussion for the potential of constraining the QS
model when combining multiple GW events from future GW observatories.  Recently,
\citet{Zhang:2020jmb} proposed a unified QS EOS including the nonzero strange
quark mass, perturbative (QCD) corrections and color superconductivity. It has
four model parameters, and we will call it the 4-dimensional (4-d) model. In
this model, by omitting the fourth and higher order terms of the strange quark
mass in the chemical potential, the EOS model has an analytical form, and it
reduces the number of free parameters from four to two.  This approximation,
which we call the 2-d model, is accurate enough for current observational
constraints, but may lead to significant biases when it is applied to fit
high-precision measurements from XG GW observations. 

In this work, in the framework of hierarchical Bayesian inference, we
investigate the potential of constraining the EOS of QSs with multiple GW events
in the XG detectors, and compare the behaviors of the 4-d model and the
approximated 2-d model in the EOS inference.  We find that the parameter
estimation (PE) biases of the 2-d model becomes significant as the number of GW
events accumulates, while the predicted $m$-$\Lambda$ relation is still
consistent with the 4-d model and the injected EOS.  Therefore, we conclude
that, in the EOS inference of QSs with the XG detector observations, the
approximated but analytical 2-d model is a good choice for a rapid inference of
the $m$-$\Lambda$ relation, while there is still a need for a full inference
using the  full (4-d) model to ensure a correct understanding of the microscope
physics.

The rest of this paper is organized as follows. In section~\ref{sec:model}, we
introduce the EOS model of QSs and the setting of our simulation.
Section~\ref{sec:inference} describes the methodology of hierarchical Bayesian
inference and the specifics to apply it to the EOS inference.  In
section~\ref{sec:results}, we present the inference results of the full and
approximated QS models.  We summarize and discuss our results in
section~\ref{sec:summary}.

%=============================
\section{QS Models and GW Signals}
\label{sec:model}
%=============================

%=============================
\subsection{EOS of QSs and Tidal Deformability} 
\label{subsec:eos model}
%=============================

SQM consists of nearly equal numbers of up (u), down (d), and strange (s)
quarks, with a small fraction of electrons to maintain charge neutrality. Since
directly solving the EOS for QSs is not feasible, various models have been
proposed to describe SQM~\cite{Miao:2021nuq}. The most popular one is the MIT
bag model with first-order corrections from the perturbative QCD, effects from
finite strange quark mass, and possible existence of color
superconductivity~\cite{Alcock:1986hz, Haensel:1986qb, Alford:2004pf}. The
thermodynamical potential density can be written as~\cite{Haensel:1986qb,
Alford:2004pf, Weissenborn:2011qu}
%%%%%%%%%%%%%%%%%%%%%%%%%%%%%%%%%%%%
\begin{equation}
\label{eq:4d}
\Omega=  \sum_{i={\rm u},\, {\rm d}, \, {\rm s}, \, e^{-}}
\Omega_i^0+\frac{3\left(1-a_4\right)}{4 \pi^2} \mu^4 +B_{\mathrm{eff}}+\frac{3
m_s^4-48 \Delta^2 \mu^2}{16 \pi^2}\,,
\end{equation}
%%%%%%%%%%%%%%%%%%%%%%%%%%%%%%%%%%%%
where $\Omega_{i}^{0}$ represents the potential for type $i$ particle described
as non-interacting fermions, $\mu=\sum_i \mu_i / 3$ is the average chemical
potential of the quark matter. We neglect the mass of u and d quarks, and denote
the mass of s quark as $m_s$. The effective bag constant $B_{\rm eff}$
accounts for the contributions from the QCD vacuum. The parameter $a_4$ is
commonly taken to be $2\alpha_{s}/\pi$ to one-loop order with $\alpha_s$
being the strong coupling constant~\cite{Fraga:2001id, Xia:2019xax}.  Note that
both $B_{\rm eff}$ and $a_4$ are effective parameters characterizing
non-perturbative effects of the strong interactions. The last term in
Eq.~(\ref{eq:4d}) is added when the SQM is in a color-flavor locked state,
with the pairing gap $\Delta$ on the order of tens to a hundred
MeV~\cite{Alford:2001zr}.  In short, in this model we use four parameters,
namely, $ \big\{B_{\rm eff}^{1/4},a_4,m_s,\Delta \big\}$, to parametrized the QS
EOS.

\citet{Zhang:2020jmb} reformulated the thermodynamic potential density in
Eq.~(\ref{eq:4d}) by neglecting the contributions from electrons and omitting
terms of order ${\cal O} \big( m_s^4 \big)$ and higher, and obtained,
%---
\begin{equation}
\label{eq:2d}
\Omega = -\frac{3}{4\pi^2} \mu^4 + \frac{3 (1-a_4)}{4\pi^2} \mu^4 -
\frac{3\Delta^2 - 3m_s^2/4}{\pi^2}\mu^2 + B_{\rm eff}.
\end{equation}
%---
Using the newly defined parameter,
%---
\begin{align}
\label{eq:beta}
\beta = \frac{3\Delta^2 - 3m_s^2 /4}{\sqrt{3 a_4}},
\end{align}
%---
\citet{Zhang:2020jmb} reparametrized the EOS using only two parameters
$\big\{B_{\rm eff}^{1/4},\beta\big\}$. This approximation is accurate enough to
describe SQM in most parameter space, especially for a small value of $m_{s}$. 

\begin{figure}
    \centering
    \includegraphics[width=12cm]{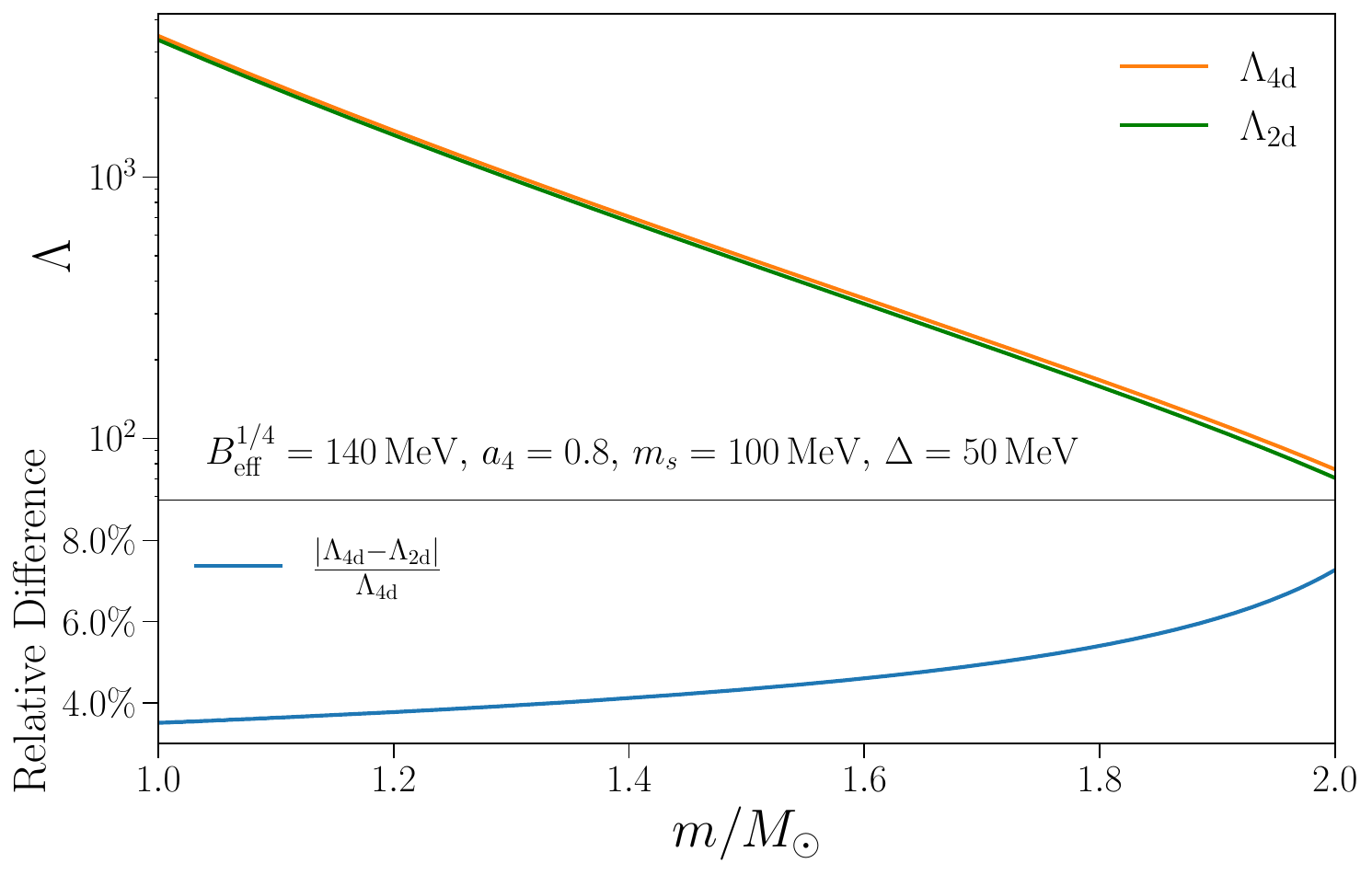}
    \caption{The upper panel shows the relation between the dimensionless tidal
    deformability $\Lambda$ and the mass $m$ for 2-d  and  4-d models. In the
    4-d model, parameters are $B_{\rm{eff}}^{1/4}=140\,\mathrm{MeV}$, $a_4=0.8$,
    $m_s=100\,\mathrm{MeV}$, and $\Delta=50\,\mathrm{MeV}$. The lower panel
    shows the relative difference in the tidal deformability between the 2-d
    and 4-d models.}
    \label{fig:mlam}
\end{figure}

The pressure,  energy density, and  baryon number density can be obtained
through the grand potential in
Eqs.~(\ref{eq:4d}--\ref{eq:2d})~\cite{Haensel:1986qb}.  By solving the
Tolman-Oppenheimer-Volkoff (TOV) equation one can obtain the mass $m$ and radius
$R$ of QSs. The tidal deformability parameter $\lambda$ of QSs/NSs is defined as
the ratio of the induced quadrupole deformation to the tidal field of the
companion~\cite{Hinderer:2007mb}. The parameter $\lambda$ can be written as 
\begin{equation}
\lambda = \frac{2}{3} k_2 R^5 \,.
\end{equation}
Here the dimensionless parameter $k_2$ is the so-called Love number, which can
be calculated from perturbation theory in the General
Relativity~\cite{Hinderer:2007mb}. The dimensionless tidal deformability is
defined as $\Lambda = \lambda/m^5$.  In figure~\ref{fig:mlam}, we present the
$m$-$\Lambda$ relation using the parameters $B_{\rm{eff}}^{1/4}=140 \,
\mathrm{MeV}$, $a_4=0.8$, $m_s=100\,\mathrm{MeV}$, and $\Delta=50
\,\mathrm{MeV}$. The corresponding parameter $\beta$ for the 2-d model can be
calculated from Eq.~(\ref{eq:beta}). In this case, the relative difference in
the tidal deformability between the 2-d and 4-d models is approximately $5\%$.

%%%%%%%%%%%%%%%%%%%%%%%%%%%%%%%%%%%%
%%%%%%%%%%%%%%%%%%%%%%%%%%%%%%%%%%%%

The Bodmer-Witten conjecture~\cite{Bodmer:1971we,Witten:1984rs} indicates that
the SQM is the ground state of strong interactions. Consequently, the energy per
baryon of SQM, $(E / A)_{\mathrm{uds}}$, must be lower than that of $^{56} \rm
Fe$, specifically $(E / A)_{\mathrm{uds}} \leqslant 930
\mathrm{MeV}$~\cite{Bodmer:1971we,Witten:1984rs}. Additionally, by imposing $(E
/ A)_{\mathrm{ud}} \geqslant 934 \mathrm{MeV}$ for the two-flavor quark matter
in its ground state, one ensures that atomic nuclei remain stable and do not
dissolve into their constituent quarks~\cite{Farhi:1984qu}. Given the specific
EOS of SQM, one obtains the mass-radius relation and the corresponding maximal
mass of QSs, $M_{\rm TOV}$. The mass measurements of massive pulsars provide a
lower bound on the maximum mass of QSs.  
Currently, the most massive pulsar
observed to date is PSR J0740+6620, with $m =
2.072^{+0.067}_{-0.066}\, M_{\odot}$ ($68\%$
confidence level)~\cite{Riley:2021pdl}. Only EOSs that support a maximum mass
$m_{\rm TOV}$ exceeding this lower bound can satisfy the observation of PSR
J0740+6620. Note that the illustrated case in figure~\ref{fig:mlam} satisfies
the above three constraints.

%=============================
\subsection{Population, Waveform and Detectors of GWs}
\label{subsec:GW models}
%=============================

This subsection describes how in the simulation we generate the GW signals,
consisting of the population model and waveform template for GW sources, and the
detector configuration for GW observations.

We adopt the mass population model of merging binaries given
by~\citet{Farrow:2019xnc} for QSs.  The distribution of the primary mass $m_1$
has two Gaussian components,
%--
\begin{equation}\label{eq:mass model}
	P(m_{1})=\gamma_{\mathrm{NS}}\mathcal{N}\left(\mu_1, \sigma_1\right) +
	\left(1-\gamma_{\mathrm{NS}}\right)\mathcal{N}\left(\mu_2, \sigma_2\right)\,,
\end{equation}
%--
where $\gamma_{\mathrm{NS}} = 0.68$, $\mu_{1} = 1.34\, M_{\odot}$, $\sigma_{1} =
0.02\,M_{\odot}$, $\mu_{2}=1.47\,M_{\odot}$, and $\sigma_{2} = 0.15\,M_{\odot}$.
Sometimes we still use the notation ``NS'', but it is clear from the context
that we are considering QSs.  The secondary mass $m_2$ is drawn from a uniform
distribution, ${\cal U}(1.14\,\ms,1.46\,\ms)$.  Considering formation and
evolution models, neither of the binary components are likely to have high
spins,  so we neglect the spin effects of the QSs, just like what
\citet{Golomb:2021tll} did for NSs.

For the extrinsic parameters of GWs, we draw the merging binaries with
isotropically distributed positions and orientations.  Using the cosmological
parameters provided by the Planck Collaboration~\cite{Planck:2018vyg}, we
generate 1000 GW events within $500 \, {\rm Mpc}$ uniformly distributed in the
comoving volume.  This corresponds to a several years' observation for a uniform
local merger rate $R = 320^{+490}_{-270}\, {\rm Gpc}^{-3}{\rm yr}^{-1}$ from the
current GW detections~\cite{LIGOScientific:2020aai}.  It is found that the major
information about the EOS is obtained from the loudest several events in the
population~\cite{Lackey:2014fwa}.  To reduce the calculation cost, we only
select the 25 sources with the highest signal-to-noise ratio (SNR) for Bayesian
analysis, leaving a more comprehensive study using the full population for
future work.  All GW signals are injected with an arbitrary merger time $t_c
=0$.

For the tidal deformability parameter, we adopt the 4-d QS model to inject the
dimensionless tidal deformability $\Lambda$ for each source.  The underlying
values of the EOS parameters are chosen to be consistent with current
observational constraints, and we use $B_{\rm eff}^{1/4} = 140\, {\rm MeV}$,
$a_4 = 0.8$, $m_s = 100\, {\rm MeV}$ and $\Delta = 50\, {\rm
MeV}$~\cite{Zhou:2017pha,Miao:2021nuq}.  In this work, we use this fixed but
typical configuration to compare the behaviors of the 4-d model and the
approximated 2-d model in the Bayesian inference of the EOS.

To include the tidal effects in the waveform, we use the {\sc imrphenomxas}
waveform model~\cite{Pratten:2020fqn,Garcia-Quiros:2020qpx,Pratten:2020ceb}, and
manually add the GW phase correction induced by the tidal deformation at the
leading order and next-to-leading order in the frequency
domain~\cite{Vines:2011ud,Favata:2013rwa,Wade:2014vqa}. The phase correction
reads
%--
\begin{equation}
    \Psi_{\rm{Tidal}}(f) = -\frac{39}{2} \tilde{\Lambda} x^5 + \left(
    -\frac{3115}{64} \tilde{\Lambda} + \frac{6595}{364} \sqrt{1 - 4\eta} \delta
    \tilde{\Lambda} \right) x^6\,,
\end{equation}
%--
where $M = m_1 + m_2$ is the total mass, $\eta = m_1 m_2 /M^2$ is the symmetric
mass ratio, and $x = (\pi G M f/c^3)^{2/3}$ is the standard post-Newtonian order
parameter.  We cut off this correction at the GW frequency corresponding to the
Schwarzschild innermost stable circular orbit~\cite{Favata:2010yd}. Note that
both the mass and the frequency are simultaneously taking values in the source
frame, or simultaneously in the detector frame, while eq.~\eqref{eq:mass model}
describes mass distribution in the source frame.  In the above equation,
$\tilde{\Lambda}$ and $\delta \tilde{\Lambda}$ are linear combinations of the
dimensionless tidal deformabilities of the two stars in the binary system,
$\Lambda_1$ and $\Lambda_2$,
%--
\begin{equation}
    \begin{aligned}
        \tilde{\Lambda} = &\  \frac{8}{13} \left[ \left(1 + 7\eta -
        31\eta^2\right)(\Lambda_1 + \Lambda_2) + \sqrt{1 - 4\eta} (1 + 9\eta -
        11\eta^2) (\Lambda_1 - \Lambda_2) \right]\,,\\
        \delta \tilde{\Lambda} = &\  \frac{1}{2} \left[ \sqrt{1 - 4\eta} \left(
        1 - \frac{13272}{1319} \eta + \frac{8944}{1319} \eta^2 \right)
        (\Lambda_1 + \Lambda_2) \right.\\
        & \quad\quad + \left. \left( 1 - \frac{15910}{1319} \eta +
        \frac{32850}{1319} \eta^2 + \frac{3380}{1319} \eta^3 \right) (\Lambda_1
        - \Lambda_2) \right].
    \end{aligned}
\end{equation}
%--
The waveform model is then given by
\begin{equation}
    \tilde{h}_{+/ \times}(f) = \tilde{h}_{+/\times}^{\rm{no\,Tidal}}(f) \cdot
    e^{-i \Psi_{\rm{Tidal}}(f)}\,,
\end{equation}
where $\tilde{h}_{+/ \times}$ is the plus/cross polarization of the GW signal in
the frequency domain, and $\tilde{h}_{+/\times}^{\rm{no\,Tidal}}$ is given by
the {\sc imrphenomxas} waveform template.  As a short summary, the waveform
model has 10 parameters, including the binary masses $m_1$ and $m_2$, the
luminosity distance $d_L$, the merger time $t_c$, the right ascension $\alpha$
and declination $\delta$, the inclination angle $\iota$, the polarization angle
$\psi$, and the two tidal deformabilities $\Lambda_1$ and $\Lambda_2$. 

For the GW detectors, we choose a ground-based network consisting of two CE
detectors and one ET detector, whose sensitivity curves are taken as
CE-2~\cite{Punturo:2010zz,Hild:2010id,Sathyaprakash:2012jk} and
ET-D~\cite{Reitze:2019iox,Reitze:2019dyk}, respectively. The two CE detectors
are positioned at the current locations of the two LIGO detectors, while
the ET detector is situated at the present Virgo detector site with a triangular shape.

%=============================
\section{Hierarchical Inference of EOS Parameters}
\label{sec:inference}
%=============================

%=============================
\subsection{Principles}
\label{subsec:principles}
%=============================

The EOS of QSs cannot be directly measured from GW signals.  Instead, different
EOSs determine different relationship between the tidal deformability and the
mass of the compact stars, leading to an impact on the GW signals.  In turn, one
firstly estimates the masses and the tidal deformabilities of the merging
binaries from GW signals, then use them to fit the $m$-$\Lambda$ relation and
estimate the EOS parameters. 

It is expected that the measurement of EOSs requires an appropriate combination
of multiple GW events.  However, if one naively combines all data and estimate
the GW parameters and EOS parameters together, say, in one giant Bayesian PE,
the inference would be computationally impractical for a high-dimensional
parameter space.  To address this issue, we adopt the so-called hierarchical
Bayesian inference method, which has been widely used for studying the
population properties of compact binaries~\cite{Mandel:2009pc, Mandel:2009nx,
Adams:2012qw, Mandel:2018mve, Thrane:2018qnx, LIGOScientific:2020aai,
KAGRA:2021duu}.  As its name suggests, the hierarchical inference allows one to
conduct inference calculation layer by layer, and estimate the GW parameters
(controlling the GW signals) and population hyper-parameters (controlling the
prior distribution of the GW parameters) separately.  In this study, the EOS
parameters can also be regarded as hyper-parameters, which give a deterministic
relationship---a $\delta$-function-type prior---between $\Lambda$ and $m$. These
hyper-parameters can be estimated in the hierarchical Bayesian inference
framework~\cite{Lackey:2014fwa, Landry:2018prl, Golomb:2021tll,
Criswell:2022ewn}.  Below we briefly introduce the methodology of the
hierarchical inference, and the procedure to apply this technology to infer  EOS
parameters.

We start from the Bayes' theorem,
%--
\begin{equation}\label{eq:bayes for hyper}
    P(\bmH|D)\propto P(D|\bmH)\pi(\bmH)\,,
\end{equation}
%--
where $D\equiv \{d_1,d_2,\cdots,d_n\}$ denotes all the collected GW data, $\bmH$
is the set of hyper-parameters.  Specifically, we have $\bmH =  \big\{B_{\rm
eff}^{1/4},a_4,m_s,\Delta \big\}$ for the 4-d model and $\bmH = \big\{B_{\rm
eff}^{1/4},\beta \big\}$ for the 2-d model.  $P(\bmH|D)$ is the posterior
distribution of the hyper-parameters, $P(D|\bmH)$ is the likelihood function,
and $\pi(\bmH)$ is the prior distribution of the hyper-parameters. As it is the
$i$-th set of GW parameters $\bmt_i$ that influence the corresponding GW data
$d_i$, the likelihood $P(D|\bmH)$ can be written as
%-- 
\begin{equation}
    \begin{aligned}
        P(D|\bmH) &= \prod_{i=1}^{n} \int P(\bmt_i|\bmH)P(d_i|\bmt_i,\bmH) \rmd \bmt_i\\
        &= \prod_{i=1}^{n} \int P({\bm m}_i|\bmH)P({\bm \Lambda}_i|{\bm
        m}_i,\bmH)P({\bm \xi}_i|{\bm m}_i,{\bm \Lambda}_i,\bmH)P(d_i|\bmt_i)
        \rmd {\bm m}_i \rmd {\bm \Lambda}_i \rmd {\bm \xi}_i\\
        &= \prod_{i=1}^{n} \int P({\bm m}_i|\bmH)\delta \Big(\bm \Lambda_i - \bm
        \lambda({\bm m}_i,\bmH)\Big)\rmd {\bm m}_i \rmd {\bm \Lambda}_i\int
        P({\bm \xi}_i)P(d_i|\bmt_i)  \rmd {\bm \xi}_i\,.
    \end{aligned}\label{eq:deconpose likelihood}
\end{equation}
%--
The first line uses the total probability theorem, under the assumption that
different GW events are independent.  In the second line, we divide the GW
parameters into three categories: the mass parameters ${\bm m}_i$, the tidal
parameters ${\bm \Lambda}_i$, and the nuisance parameters ${\bm \xi}_i$.  Note
that one can choose different but equivalent representations for mass
parameters, namely $\{m_{1,i},m_{2,i}\}$ versus $\{{\cal M}_i,\eta_i\}$, and
tidal parameters, namely $\{\Lambda_{1,i},\Lambda_{2,i}\}$ versus
$\{\tilde{\Lambda}_i,\delta \tilde{\Lambda}_i\}$.  In the third line, the
$\delta$-function means that ${\bm \Lambda}_i$ is determined by ${\bm m}_i$ and
$\bmH$ through the EOS model $\bm \lambda({\bm m}_i,\bmH)$, which is given in
section \ref{subsec:eos model}.  Besides, we assume that the distributions of
nuisance parameters ${\bm \xi}_i$ are independent of ${\bm m}_i$, ${\bm
\Lambda}_i$ and $\bmH$.

In the hierarchical inference, the integral of the nuisance parameters, $\int
P({\bm \xi}_i)P(d_i|\bmt_i)  \rmd {\bm \xi}_i$, is independent of the
hyper-parameters $\bmH$, thus can be calculated in advance to reduce the
computational cost~\cite{Lackey:2014fwa,Thrane:2018qnx,Golomb:2021tll}. 
Considering the single-event PE for the $i$-th GW event, the posterior of
parameters $\bmt_i$ can be written as 
%--
\begin{equation}\label{eq:single PE}
    P(\bmt_i|d_i) \propto \pi_\varnothing(\bmt_i) P(d_i|\bmt_i)\,,
\end{equation}
%--
where $P(d_i|\bmt_i)$ is the single-event likelihood~\cite{Finn:1992wt},
%--
\begin{equation}
    L_i(\bmt_i) \equiv P(d_i|\bmt_i) \propto
    e^{-\frac{1}{2}\big(d_i-h(\bmt_i),d_i-h(\bmt_i)\big)}\,,
    \label{eq:single likelihood}
\end{equation}
%--
with $h(\bmt_i)$ given by the waveform model in section~\ref{subsec:GW models}.
In eq.~(\ref{eq:single likelihood}) the inner product $(u,v)$ is defined as
%--
\begin{equation}
    (u,v): = 2\Re\int_{-\infty}^{\infty}\frac{u^{*}(f)v(f)}{S_n(|f|)} { \rmd} f,
\end{equation}
%--
where $S_n$ is the power spectral density of the noise of the detector network,
$u(f)$ and $v(f)$ are the Fourier transforms of $u(t)$ and $v(t)$, respectively.
A more detailed description about the single-event likelihood, such as the
projection of GWs and combination of different detectors, can be found in
ref.~\cite{Lackey:2014fwa}.  Now we focus on the $\pi_\varnothing(\bmt_i)$ in
eq.~(\ref{eq:single PE}), which is some {\it default} prior of $\bmt_i$ chosen
for a complete expression of the Bayes' theorem.  However, if one chooses
$\pi_\varnothing(\bmt_i) \propto P({\bm \xi}_i)$, namely, flat priors for mass
and tidal parameters and the same marginal distributions in
eq.~\eqref{eq:deconpose likelihood} for nuisance parameters,  the joint
posterior distribution of ${\bm m}_i$ and ${\bm \Lambda}_i$ will become 
%--
\begin{equation}
    P({\bm m}_i,{\bm \Lambda}_i|d_i) = \int P(\bmt_i|d_i) \rmd {\bm \xi}_i
    \propto \int P({\bm \xi}_i)P(d_i|\bmt_i)  \rmd {\bm \xi}_i\,,
\end{equation}
%--
which is just the integral over $P({\bm \xi}_i)$ in eq.~\eqref{eq:deconpose
likelihood}, modulus a constant multiplicative factor.

From above derivation, one finds that eq.~\eqref{eq:single PE} represents an
auxiliary PE, whose posterior can help to calculate the integral on the nuisance
parameters in eq.~\eqref{eq:deconpose likelihood}.  In the language of Bayesian
analysis, this integral is called the marginal
likelihood~\cite{Loredo:2024bayesian}, or the
quasi-likelihood~\cite{Lackey:2014fwa}, which is denoted as $L^{\rm q}$,
%--
\begin{equation}
    L^{\rm q}_i({\bm m}_i,{\bm \Lambda}_i) = \int P({\bm \xi}_i)L_i({\bm
    m}_i,{\bm \Lambda}_i,{\bm \xi}_i) \rmd {\bm \xi}_i \propto P({\bm m}_i,{\bm
    \Lambda}_i|d_i)\,.
\end{equation}
%--
Furthermore, the calculation of posterior in eq.\eqref{eq:single PE} is a
standard single-event PE, and independent of the hyper-parameters $\bmH$. Thus, 
one can firstly conduct these auxiliary PEs for each GW event by some sampling
techniques, such as the Markov-Chain Monte Carlo (MCMC)
method~\cite{Christensen:1998gf, Christensen:2004jm, Sharma:2017wfu} and nested
sampling~\cite{2004AIPC..735..395S, Skilling:2006gxv}, and generate samples of
the posterior distribution $P(\bmt_i|d_i)$. Then, the quasi-likelihood $L^{\rm
q}_i({\bm m}_i,{\bm \Lambda}_i)$ can be obtained by some density-estimate
methods, such as the kernel density estimates, Gaussian processes, and Gaussian
mixture models~\cite{Rosenblatt:1956kde, edward2006rasmussen, Lackey:2014fwa,
Landry:2018prl, Wysocki:2020myz, DEmilio:2021laf, Golomb:2021tll}.  Finally,
substituting $L^{\rm q}_i$ into eq.~\eqref{eq:deconpose likelihood} and
completing the integral involving the $\delta$-function, the posterior for
hyper-parameters becomes,
%--
\begin{equation}
    P(\bmH|D) \propto \pi({\bm H})\prod_{i=1}^{n} \int P({\bm m}_i|\bmH)L^{\rm
    q}_i \Big({\bm m}_i,{\bm \lambda}\big({\bm m}_i,\bmH)\Big) \rmd {\bm m}_i
    \,,
\end{equation}
%--
which avoids dealing with nuisance parameters and hyper-parameters
simultaneously, and makes the computation cost much reduced. 

%=============================
\subsection{Implementation}
\label{subsec:implementation}
%=============================

When conducting the single-event PE, we choose ${\cal M}$ and $\eta$ as the mass
parameters, and $\tilde{\Lambda}$ and $\delta \tilde{\Lambda}$ as the tidal
parameters. Namely, all the parameters we want to recover for a single GW event
are $\bmt = \big\{{\cal M},\eta,\tilde{\Lambda},\delta
\tilde{\Lambda},d_{L},t_{c},\alpha,\delta,\psi,\iota \big\}$. The priors of
${\cal M} $, $\eta$, $\tilde{\Lambda}$, $\delta \tilde{\Lambda}$ and $t_c$ are
uniform.  The prior of $d_L$ is uniform in comoving volume and source frame time.  We
adopt the isotropic distribution for the priors of $\alpha$, $\delta$, $\psi$
and $\iota$.  Following \citet{Lackey:2014fwa}, we use the Gaussian kernel
density estimator to calculate the quasi-likelihood.  We also treat $\delta
\tilde{\Lambda}$ as a nuisance parameter, since it contributes to the waveform
at a higher order than $\tilde{\Lambda}$~\cite{Vines:2011ud, Favata:2013rwa,
Wade:2014vqa}, and therefore it is much less informative for the EOS
parameters~\cite{Lackey:2014fwa}. Both choices are studied in our work.

%--
\begin{table}[tbp]
    \def\arraystretch{1.4}
\tabcolsep=0.3cm
    \centering
    \begin{tabular}{|clcc|c|}
    \hline
    Model & Parameter &True Value & Prior &Posterior\\
    \hline
     4-d&$B_{\rm eff}^{1/4}$ $(\rm{MeV})$ & 140&${\cal U}(130,150)$ & $139.0^{+1.7}_{-1.4}$\\
    &$a_4$ &0.8& ${\cal U}(0.4,1.0)$ & $0.7^{+0.2}_{-0.2}$\\
    &$m_s$ $(\rm{MeV})$ & 100&${\cal U}(75,125)$ & $96.8^{+18.1}_{-15.1}$\\
    &$\Delta$ $(\rm{MeV})$ & 50&${\cal U}(0,100)$ & $41.2^{+14.4}_{-19.6}$\\
    \hline
    2-d &$B_{\rm eff}^{1/4}$ $(\rm{MeV})$ & 140&Flat & $138.2^{+1.5}_{-1.6}$\\
     &&140& Transformed & $137.8^{+1.6}_{-1.3}$\\
&$\beta$ $(\rm{MeV}^2)$ & 0&Flat & $-1635^{+2297}_{-2229}$\\
     &&0& Transformed & $-2415^{+2312}_{-1751}$\\
    \hline
    \end{tabular}
    \caption{The injected true values and priors of the EOS parameters in 4-d
    and 2-d models for the Bayesian inference. The corresponding medians and
    central 68\% credible intervals of the posteriors are  listed in the last
    column. Note that the data used in all inferences are generated with the 4-d
    model.}\label{tab:prior} 
\end{table}
%-- 

For the EOS parameters, we choose flat priors for $\big\{B_{\rm
eff}^{1/4},a_4,m_s,\Delta \big\}$ around the injected values. While for
$\big\{B_{\rm eff}^{1/4},\beta \big\}$ in the 2-d model, the prior of $\beta$
has two choices: (i) it can be chosen to be flat in the allowed range; (ii) it
can be  generated by the priors of the 4-d model according to the probability
density transformation. The latter choice is to ensure a consistent comparison
between the 2-d and 4-d models, while the former choice suits for testing the
unbiasedness of the 2-d model and avoiding priors' influence. We summarize the
priors of the EOS parameters in table~\ref{tab:prior}. 

In PEs of both the single-event parameters and the hyper-parameters, we generate
the posterior samples using the {\sc Bilby} implementation~\cite{Ashton:2018jfp}
of {\sc Nessai}~\cite{nessai,Williams:2021qyt,Williams:2023ppp}.  To investigate
the 4-d model performance in the EOS inference, we conduct two PEs with the
loudest 25 GW events generated in section~\ref{subsec:GW models}. The first PE
uses the flat priors in table~\ref{tab:prior} and focuses on the EOS's
dependence on different parameters. The second one includes some observational
constraints for QSs, giving a more comprehensive forecast for measuring the EOS
parameters with the XG GW detectors. 

%=============================
\section{Results and Discussions}
\label{sec:results}
%=============================

%=============================
\subsection{Inference of Parameters in the 4-d Model}
\label{subsec:4d results}
%=============================

%--
\begin{figure}[tbp]
    \centering \hspace{-0.5cm}
    \includegraphics[width=0.5\textwidth]{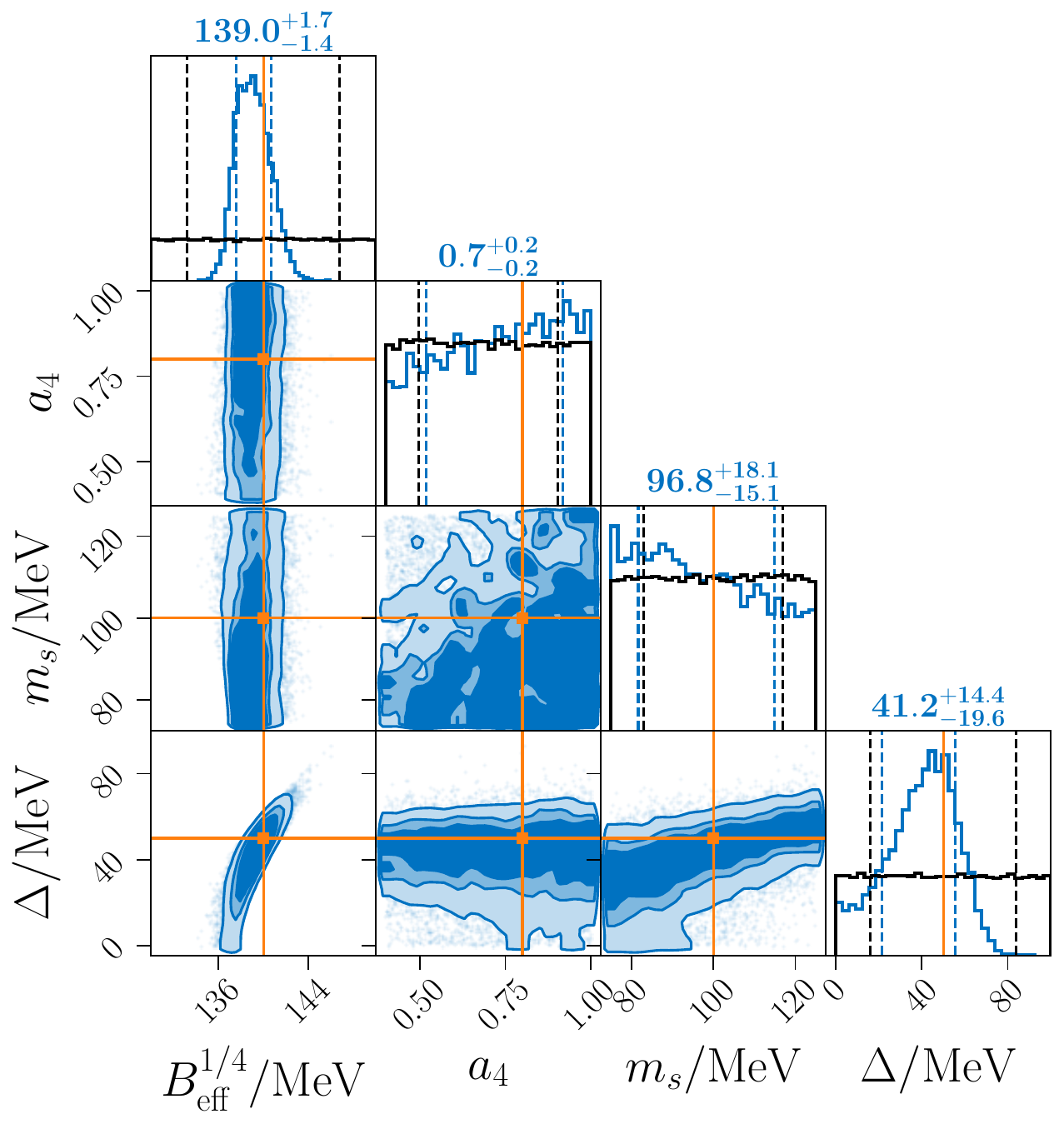}
    \includegraphics[width=0.5\textwidth]{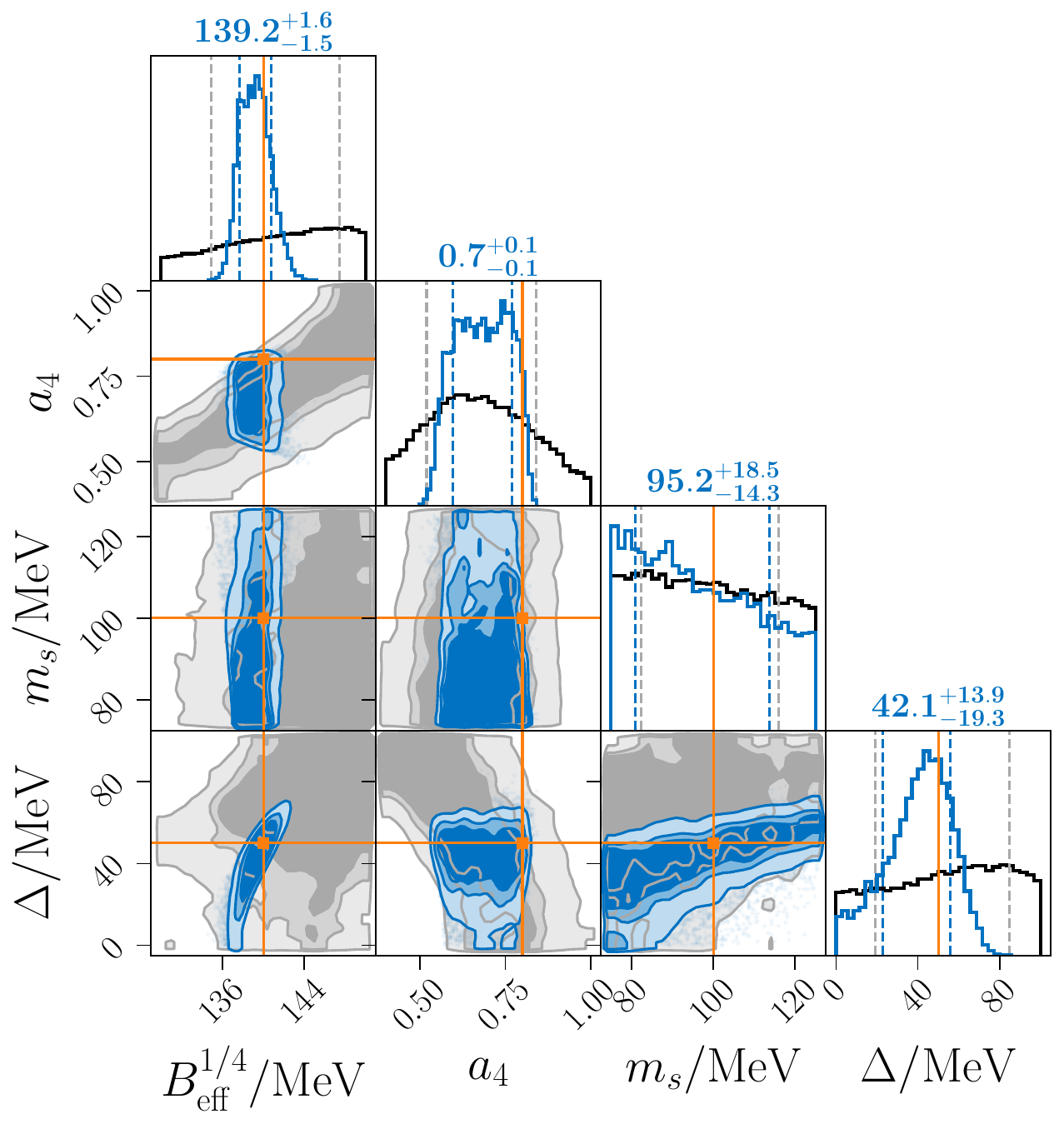}
    \caption{Posterior distributions of the EOS parameters in the 4-d model. The
    contours show 50\%, 68\% and 90\% credible regions. The numbers on the
    histogram represent the median and the central 68\% credible interval of
    each marginalized distribution. The orange solid line indicates the injected
    value of EOS parameters. The left panel is for flat priors of EOS
    parameters, while the right panel further requires stability conditions (see text),
    which are given by the grey areas, and the constraint from  PSR
    J0740+6620~\cite{NANOGrav:2019jur, Riley:2021pdl}. 
    }\label{fig:4d posterior} 
\end{figure}
%--

We summarize the inference results in figure~\ref{fig:4d posterior} for the 4-d
model.  In the left panel of figure~\ref{fig:4d posterior}, we show the
posterior distribution corresponding to the flat priors.  We also show the
marginalized posterior distributions, together with their median and the central
68\% credible intervals of each parameter in the diagonal.  The injected values
of the EOS parameters are recovered well, falling into the 68\% credible
intervals.  For $B_{\rm eff}^{1/4}$ and $\Delta$, the median seems to slightly
deviate from the injected value.  However, note that the mode of the posterior
distribution is just the maximum likelihood estimation (MLE) of the parameters
when flat priors are chosen, and the MLE is not necessarily coincident with the
median of each parameter.  We check that the posterior mode agrees well with the
injected values, which can also be observed intuitively by the joint
distribution of $B_{\rm eff}^{1/4}$ and $\Delta$ in the lower left  corner.  The
parameters $a_4$ and $m_s$ are not well constrained, whose posterior
distributions are broad and flat, resembling the priors plotted with black
lines.  Therefore, the median is also dominated by the priors, which was also
observed in refs.~\cite{Li:2020wbw,Miao:2021nuq}.  This result is consistent
with the fact that $a_4$ and $m_s$ only have a weak impact on the $m$-$\Lambda$
relation~\cite{Alford:2004pf,Zhou:2017pha,Li:2020wbw,Miao:2021nuq}.  Besides, We
find that $m_s$ has a slight degeneracy with $\Delta$. This originates from the
EOS's dependence on the quadratic coefficient $a_2 =
m_s^2-4\Delta^2$~\cite{Alford:2004pf}, which is also the numerator of $\beta$ in
the 2-d model. 

When calculating the posterior in the right  panel, we further add the stability
requirements for QSs in the EOS model: (i) the normal atomic nuclei are more
stable than non-strange quark matter; (ii) the SQM is expected to be more stable
than normal nuclear matter~\cite{Bodmer:1971we,Witten:1984rs}.  These two
constraints are called the ``2 flavor'' line and the ``3 flavor'' line
respectively~\cite{Weissenborn:2011qu}, and single out a subset in the priors'
parameter space.  We plot this constrained region in the corner plot with grey
contours, and show the marginalized priors with black lines.  Besides, the
maximum QS mass of an EOS needs to be large enough to support the known heaviest
pulsars.  We adopt the mass measurement of PSR
J0740+6620~\cite{NANOGrav:2019jur, Riley:2021pdl}, $m =
2.072^{+0.067}_{-0.066}\, M_{\odot}$.  With these requirements, the quartic
coefficient $a_4$ is constrained more tightly than in the case of flat priors.
This is  because that the stability requirements introduce some mild
correlations between $a_4$ and the well-constrained parameters $B_{\rm
eff}^{1/4}$ and $\Delta$. 

Thanks to the high sensitivity of XG GW detectors, the measurement precision of
$B_{\rm eff}^{1/4}$ and $\Delta$ can be improved by several times compared to
the constraints from GW170817 and GW190425 in current-generation detector
network~\cite{Zhou:2017pha,Miao:2021nuq}.  For example, the color-flavor locked
pairing gap $\Delta$ is barely measurable with the current GW detectors, and the
reported constraints highly depend on the choices of priors~\cite{Miao:2021nuq}.
As a comparison, one can find a significant peak in the marginalized posterior
distribution of $\Delta$ in figure~\ref{fig:4d posterior}. These constraints
will be further tightened if more lower-SNR (besides the top 25) events are
added into the analysis.

%=============================
\subsection{Comparison between 2-d and 4-d Models}
\label{subsec:comparison}
%=============================

In the 4-d model, we find that $a_4$ and $m_s$ are not constrained well, and
there exists a weak correlation between $m_s$ and $\Delta$. This implies
possible redundance in the parameterization of the EOS model. As mentioned in
section~\ref{subsec:eos model}, \citet{Zhang:2020jmb} proposed a QS model with
only 2 parameters, $B_{\rm eff}^{1/4}$ and $\beta$, reducing the number of
degrees of freedom. However, the simplification is based on omitting the fourth
and higher-order terms of $m_s$ in the chemical potential, which may lead to
biases in the inference of EOS parameters. In this subsection, we analyze the
EOS inferences using the 2-d model and test whether the EOS parameters are
recovered correctly.

%--
\begin{figure}[tbp]
    \centering
    \includegraphics[width=0.45\textwidth]{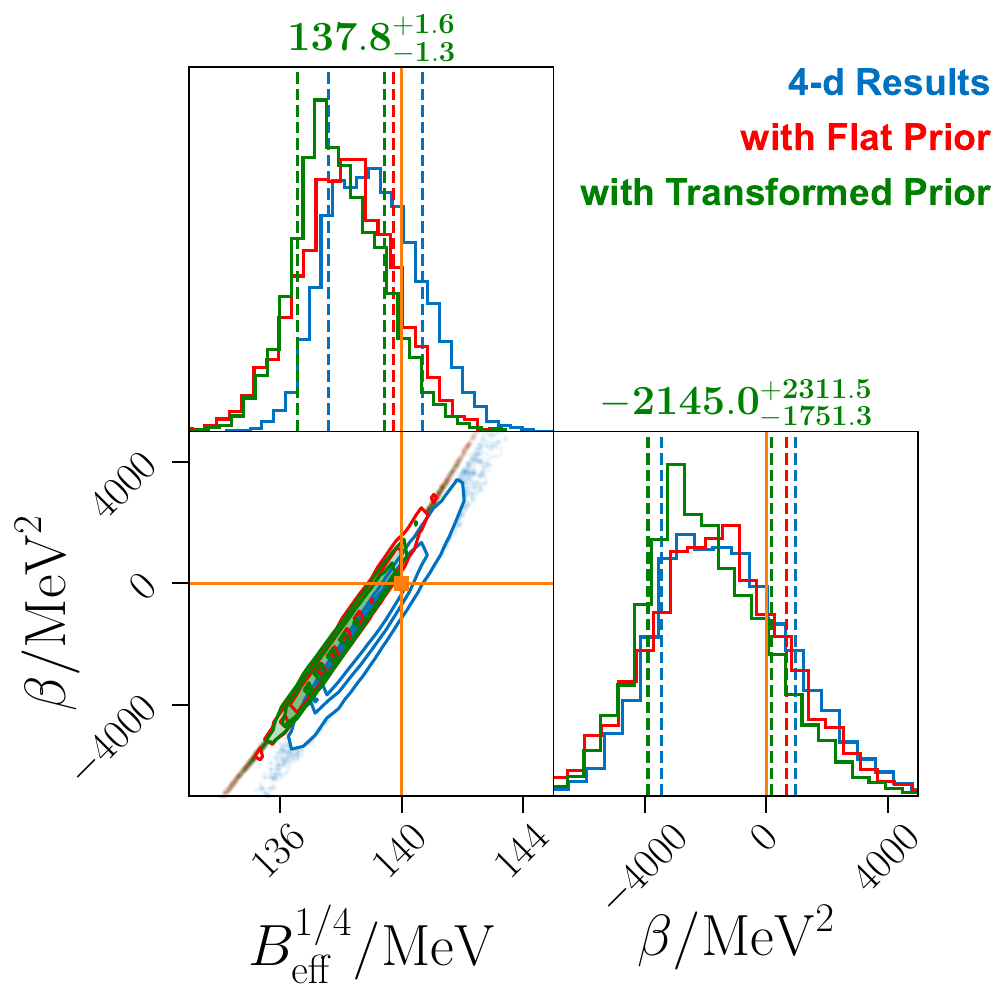} \hspace{0.5cm}
    \includegraphics[width=0.45\textwidth]{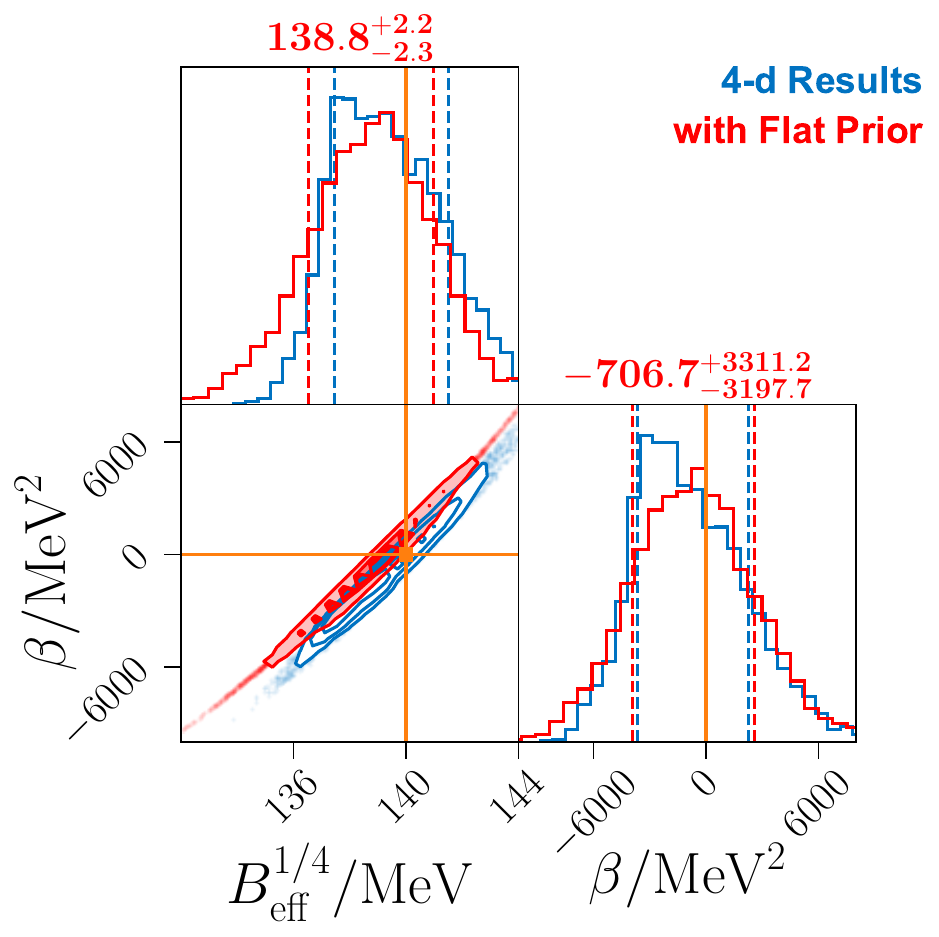}
    \caption{Similar to figure~\ref{fig:4d posterior}, but for the 2-d model.
    The blue contours are drawn according the transformed samples from the 4-d
    posteriors without observational constraints (i.e., the left panel of
    figure~\ref{fig:4d posterior}). The red and green contours represent the
    inference results directly using the 2-d model with two different priors
    (see table~\ref{tab:prior}). The left panel shows posterior using the 25
    loudest GW events in the population, while in the right panel only 4 loudest
    GW events are used. } 
    \label{fig:2d posterior}
\end{figure}
%--

Like the 4-d model, we show the posterior distributions of $B_{\rm eff}^{1/4}$
and $\beta$ in figure~\ref{fig:2d posterior}.  In the left panel, the red and
green contours respectively represent the inference results using the flat and
transformed priors in table~\ref{tab:prior}.  One immediately finds that the
injected values are not in the 90\% credible region of the posterior
distribution. For the marginal distributions, the injected value of $B_{\rm
eff}^{1/4}$ is also beyond the 68\% credible interval, while the injected value
of $\beta$ is close to the upper bound of the 68\% credible interval.  As a
comparison, the 4-d posterior samples in section~\ref{subsec:4d results} are
transformed into the 2-d parameter space and they are plotted in blue.  We find
that the injected values still fall into the contours in this transformed space.
Therefore, we conclude that using the 2-d simplified QS model may introduce
systematic biases in the EOS inference with the 25 loudest GW events in the XG
GW detectors.  It should be emphasized that obvious biases are arising because
of the accumulation of high-precision GW events.  In the right panel of
figure~\ref{fig:2d posterior}, we show the inference results using 2-d and 4-d
models with only the 4 loudest GW events.  In this case, the statistical
uncertainties of the EOS parameters are large enough to hide the biases. 

%--
\begin{figure}[tbp]
    \centering
    \includegraphics[width=0.7\textwidth]{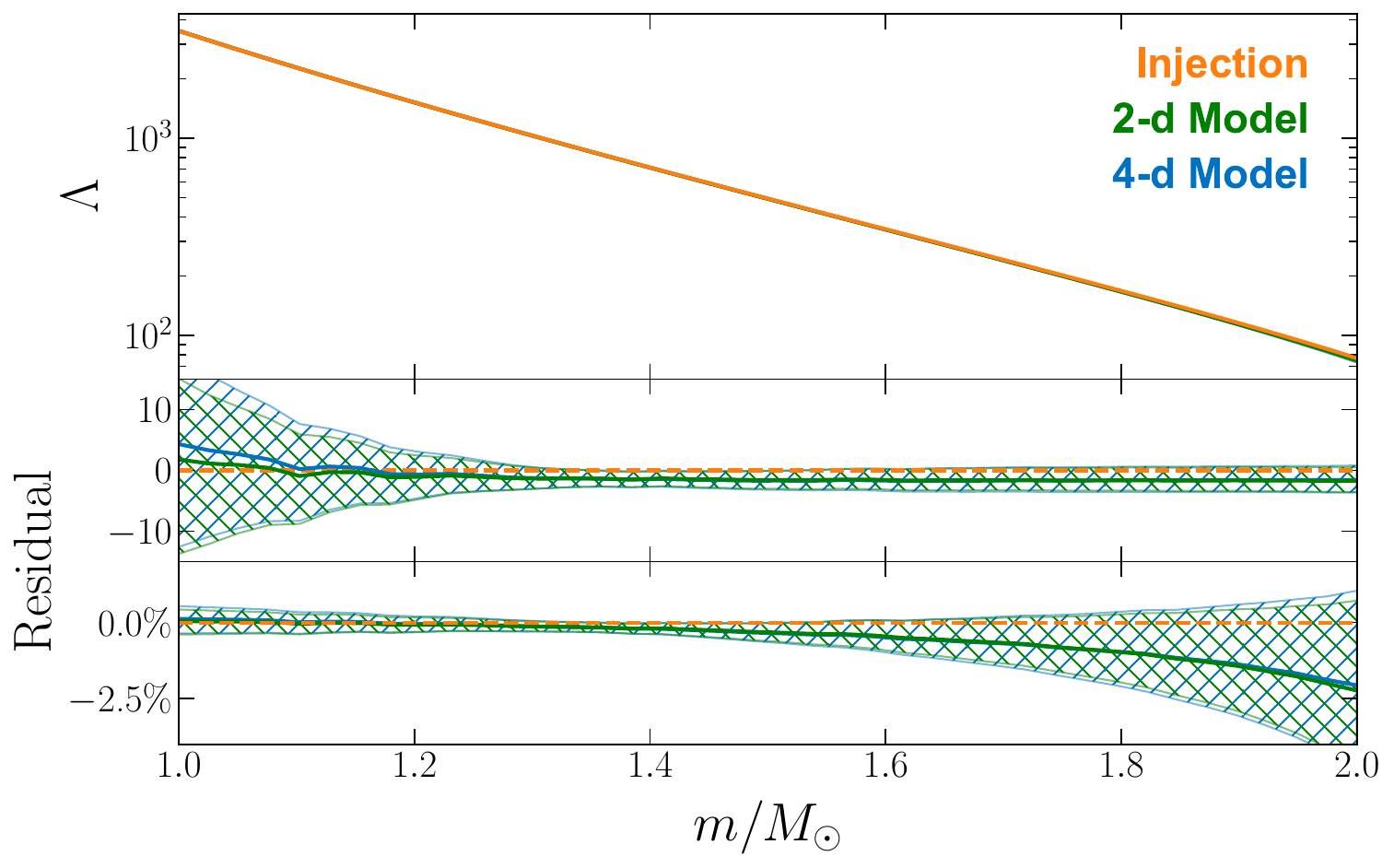}
    \caption{The $m$-$\Lambda$ relations for the posterior of the 2-d and 4-d
    EOS models, plotted in green and blue respectively. The orange line
    represents the injected $m$-$\Lambda$ relation. For each QS mass, we show
    the median and the 68\% credible interval of $\Lambda$.}\label{fig:lambda-m}
\end{figure}
%--

We plot the $m$-$\Lambda$ relation in figure~\ref{fig:lambda-m} according to the
posterior samples of the 2-d and 4-d models. We choose the posterior samples of
the 4-d model without additional constraints (i.e., posteriors in the left panel
of figure~\ref{fig:4d posterior}), while for the 2-d model, we use the posterior
samples with transformed priors to ensure a consistent comparison. Even through
the inference of the 2-d model has obvious biases in the EOS parameters, the
recovered $m$-$\Lambda$ relation is very close to that of the 4-d model. The
recovered $m$-$\Lambda$ relationships in both cases are consistent with the
injected EOS within the 68\% credible level. This is not surprising, since in
this work the EOS parameters are constrained by measuring $\Lambda$ and $m$ from
the GW events. Furthermore, we find a strong degeneracy between $B_{\rm
eff}^{1/4}$ and $\beta$ in the inference of the 2-d model. This may imply that
there still exists redundant degrees of freedom in the 2-d QS model, but it is
more likely due to the fact that only the tidal parameter of the QS is used to
constrain the EOS. Similar degeneracy is also observed when transforming the 4-d
posterior samples to the 2-d parameter space. We leave a more detailed study of
this degeneracy for future work.

%=============================
\section{Summary and Outlook}
\label{sec:summary}
%=============================

In this work, we simulated GW observations with an XG GW detector network
consisting of two CE detectors and one ET detector, and tested the potential of
constraining the EOS of QSs with these GW events.  Adopting the hierarchical
Bayesian inference method~\cite{Adams:2012qw,Lackey:2014fwa,Mandel:2018mve}, we
compared the inference results of the full 4-d model~\cite{Alford:2004pf} and the
simplified 2-d model~\cite{Zhang:2020jmb}. 

From the  inference of the 4-d model in figure~\ref{fig:4d posterior}, we found
that the constraints on the effective bag constant $B_{\rm eff}^{1/4}$ and the
pairing energy gap of the color superconductivity $\Delta$ can be improved by
several times with the help of the XG detectors, while the strange quark mass
$m_s$ and the QCD correction parameter $a_4$ are still poorly constrained. 
Adding the stability and maximum-mass requirements does not improve the
constraints on $B_{\rm eff}^{1/4}$ and $\Delta$, but slightly tightens the
constraint on $a_4$ because of the mild correlations between $a_4$ and the
well-constrained parameters.  Therefore, $m_s$ and $a_4$ are not likely to be
directly inferred from the tidal deformation measurements of GWs in XG
detectors.  These results are consistent with the previous
studies~\cite{Alford:2004pf, Zhou:2017pha, Li:2020wbw, Miao:2021nuq}.

Figure~\ref{fig:2d posterior} showed the inference results of the 2-d and 4-d
models, and demonstrated how the PE biases arise with the accumulation of GW
events in the 2-d model.  When taking the 25 loudest GW events into the
analysis, the inference results of the 2-d model show systematic biases in the
EOS parameters, while the 4-d model gives unbiased results.  If only the top 4
loudest events are used, the statistical uncertainties of the EOS parameters are
large enough to cover the biases of the 2-d model.  However, the predicted
$m$-$\Lambda$ relations in both the 2-d and 4-d models are consistent with the
injected relation within the 68\% credible level, even when 25 loudest GW events
are used.  Therefore, if one only focuses on the tidal parameter of QSs, the 2-d
model is suitable as a simplified and analytical expression of the EOS,
providing a smaller parameter space thus faster computation.  At the same time,
we emphasized that the approximate accuracy of the 2-d model is insufficient to
meet the observation precision in the XG GW detections, especially when
analyzing large numbers of GW events.

To our knowledge, this study is the first to discuss the hierarchical inference
of the QS EOS with a large number of GW events in the XG GW detectors.  However,
there are still some limitations in this work.  First, to obtain an unbiased
result for the EOS parameters, the single-event PE should also be unbiased. 
Some factors, such as the inaccurate waveform model, may lead to non-negligible
biases in the measurements of the GW parameters, especially for  high-SNR
events~\cite{Cutler:2007mi,Williamson:2017evr,Purrer:2019jcp,Gamba:2020wgg}.
Besides, considering the high detection rate in the XG detectors, the GW signals
may overlap with each other, which can also lead to PE
biases~\cite{Pizzati:2021apa,Hu:2022bji,Wang:2023ldq}.  Moreover, the spins of
the compact stars were ignored in this work.  This may lead to an overestimation
of the measurement precision of the tidal and mass parameters, thus obtain an
optimistic constraint on the EOS parameters. There are also some novel
waveform model characterizing the tidal effects more accurately, such as
the NRTidalv1--NRTidalv3 models~\cite{Dietrich:2017aum, Dietrich:2019kaq, Abac:2023ujg}. These models include more physical
effects compared to post-Newtonian corrections, and may help to break the degeneracies in the inference of the
tidal parameters.
 As for the inference at the
hyper-parameter level, it is natural to study the inference of the EOS
parameters using the whole population. Using GPU acceleration can significantly
reduce the computational time of the hierarchical inference including a large
number of GW events, and it has been applied to the study of
NSs~\cite{Talbot:2020oeu}.  Some studies also recommended inferring the EOS and
the mass distribution simultaneously to avoid possible
biases~\cite{Wysocki:2020myz,Golomb:2021tll}. Finally, we found a tight
degeneracy between $B_{\rm eff}^{1/4}$ and $\beta$ in the 2-d  QS model.  One
may combine other information, such as the mass and radius observation or the
post-merger remnants to break this degeneracy. The nonradial oscillation of QSs, such as the $f$ and the $w_{\rm II}$ modes, can also leave imprints in GW signals and serve as a probe for EOS inference~\cite{Sotani:2003bg, Li:2022qql, Li:2024hzt}. We leave these extensions for
future work.  In conclusion, the XG GW observation has a great potential to
constrain the EOS of compact stars, and future observations will further uncover
the properties of dense matter.

%=============================
\acknowledgments

This work was supported by the National Natural Science Foundation of China
(123B2043, 11991053, 12405065, 12405052), the Beijing Natural Science Foundation
(1242018), the National SKA Program of China (2020SKA0120300), the Max Planck
Partner Group Program funded by the Max Planck Society, the Fundamental Research
Funds for the Central Universities and the High-performance Computing Platform
of Peking University. 

\bibliographystyle{apsrev4-1}
\bibliography{ref}
\end{document}